\documentclass[aps,prl,twocolumn,showkeys,showpacs,superscriptaddress]{revtex4-1}
\usepackage{amssymb,amsmath}
\usepackage{graphicx}
\usepackage{hyperref}
\usepackage[font=small,labelfont=bf,
   justification=justified,
   format=plain]{caption}
\usepackage{subcaption}

\newcommand\be{\begin{equation}}
\newcommand\ba{\begin{eqnarray}}
\newcommand\ee{\end{equation}}
\newcommand\ea{\end{eqnarray}}

\begin{document}


\title{Inflation and the Measurement Problem}

\author{Stephon Alexander}
\email{Stephon\_Alexander@brown.edu}
\affiliation{Department of Physics, Brown University, Providence, RI, 02912, USA}
\author{Dhrubo Jyoti}
\email{Dhrubo.Jyoti.gr@dartmouth.edu}
\affiliation{Center for Cosmic Origins, Wilder Laboratory, Dartmouth College, Hanover, NH, USA}
\author{Jo\~{a}o Magueijo}
\email{j.magueijo@imperial.ac.uk}
\affiliation{Theoretical Physics, Blackett Laboratory, Imperial College, London, SW7 2BZ, United Kingdom}

\date{\today}

\begin{abstract}\noindent
We propose a solution to the quantum measurement problem in Inflation. Our model treats Fourier modes of cosmological perturbations as analogous to particles in a weakly-interacting Bose gas. We generalize the idea of a macroscopic wavefunction to cosmological fields, and construct a self-interaction Hamiltonian that focuses that wavefunction. By appropriately setting the coupling between modes, we obtain the standard adiabatic, scale-invariant power spectrum. Because of Central Limit Theorem (CLT), we recover a Gaussian Random Field, consistent with observations.  \end{abstract} 

\maketitle
\noindent {\bf Introduction -- } Inflation is a very successful paradigm, solving the Horizon, Flatness and Monopole problems. But perhaps its most interesting aspect is that, it traces the origin of structure in the Universe to quantum zero-point fluctuations \cite{Starobinsky:1980te,Guth:1985ya}. We believe that the Universe had a quantum mechanical beginning, but how exactly did the classical universe we are familiar with emerge? While the mechanism has been studied for decades \cite{Albrecht:1992kf, Polarski:1995jg,Kiefer:2008ku}, a number of authors have pointed out important gaps in our understanding \cite{weinberg2008,padmanabhan:1996,lyth2009primordial} (for a review, see \cite{Perez:2005gh}). Proponents of the mechanism assume a ``quantum non-demolition measurement'' \cite{pointer}, and acknowledge that the current description is only ``pragmatic'' and needs eventually to be fully justified \cite{Kiefer:2008ku}.

We propose a solution to this cosmological quantum measurement problem  \eqref{CMP}. Our approach is an effective wavefunction collapse mechanism arising from a novel interaction between Fourier modes, inspired by a two-dimensional weakly-interacting Bose gas, to be contrasted with fundamental modifications to the Schr\"{o}dinger equation \cite{Martin:2012pea,Canate:2012ua,Leon:2015hwa}. Our mechanism is not a replacement but rather an add-on to the standard description. An alternative approach to the problem is Bohmian mechanics, which interprets the wavefunction as an actual field and avoids the notion of an observer collapsing the wavefunction \cite{PintoNeto:2011ui}.

The CMB has an average temperature of 2.7 K, but has small variations of order one part in $10^5$. These are signatures of slight variations in the gravitational field in different regions of the Universe at the surface of last scattering (Sachs-Wolfe effect). This primordial curvature perturbation field $\zeta(x)$, which eventually lead to the formation of large scale structures (LSS) such as galaxies, is analyzed as follows \cite{Martin:2012pea},
\be \label{alm}
a_{lm}=\frac{1}{(2\pi)^{3/2}}\int d\Omega_{\bf e}~d{\bf k} ~\frac{1}{5}~\zeta_{\bf k}~Y_{lm}^*({\bf e}) ~ e^{-i{\bf k.e}}
\ee
where we defined the Fourier modes $\zeta_{\bf k}(\eta)\equiv\frac{1}{(2\pi)^{3/2}}\int d^3x~\zeta({\bf x},\eta)~e^{i\bf k\cdot x}$~, where ${\bf k}$ is the wave-vector co-moving with expansion of space (working in natural units, $\hbar=c=k_{B}=1$). For a given $l$, the $a_{lm}$'s fit a normal distribution with mean zero and standard deviation $\sqrt{C_l}$. The standard deviation is independent of $m$, dubbed statistical isotropy of the Universe. 

The $a_{lm}$'s are essentially a weighted sum over $\zeta_{\bf k}$'s. But if the latter are independently-distributed random variables, then CLT states that, as long as the standard deviation of each variable $\zeta_{\bf k}$ is finite, the probability distribution for each $a_{lm}$ will approach a normal distribution in the limit of large number of $\zeta_{\bf k}$'s. Interestingly, this means that each $\zeta_{\bf k}$ can be drawn from {\it any} distribution. It need not be normally distributed as it is in the standard description, originating from the ground-state Gaussian wavefunction of the harmonic oscillator. In other words, because of CLT, classical gaussianity of $a_{lm}$'s does {\it not} imply quantum gaussianity of $\zeta_{\bf k}$'s; it is not an if-and-only-if relationship. CLT essentially washes out the underlying distribution, and generically yields a Gaussian Random Field. This is in concord with observations since non-Gaussianity appears to be small \cite{Ade:2013ydc}. We will utilize this flexibility in our solution to the cosmological measurement problem.
\\
\\
\noindent {\small \bf Quantum Fluctuations: Standard Description -- \normalsize}Inflation is said to generate adiabatic perturbations $\zeta({\bf x})$  as follows. This is a standard calculation  \cite{Guth:1985ya,Martin:2012pea,Dodelson}. Consider a massless scalar field minimally coupled to gravity, $S=\int d^4x \sqrt{-g}~\frac{1}{2}\left[{M_p}^2R-\partial_\mu\phi\ \partial^\mu\phi\right]$. The generalization to any single-field, slow-roll model of inflation with effective potential $V(\phi)$ is lengthy but straightforward. Inflation can be characterized by the scale factor $a(\eta)=\frac{-1}{H\eta}$, with conformal time $\eta\in[\eta_0,0)$ and Hubble parameter $H$. Perturbing the action to second order, we obtain in conformal coordinates $g_{\mu\nu}=a(\eta)^2\,diag(-1,1,1,1)$, (originally worked out in \cite{MBV}), $\delta^{(2)}S=\frac{1}{2}\int d\eta\,d^3x\left[\left(\frac{\partial v}{\partial\eta}\right)^2-\delta^{ij}\,\partial_iv\,\partial_jv+\frac{2}{\eta^2}v^2\right]$ where $v$ is the Mukhanov-Sasaki variable \footnote{$\zeta$ is the canonical total scalar perturbation in the ADM formulation of General Relativity \cite{MBV}}, 
\be\label{dynamical}
v\equiv a(\eta)\,\zeta\,.
\ee
Using the Fourier transform defined earlier, we recover the Hamiltonian
\be\label{Ham}
\mathbb{H}=\int d^3k\left[p_{\bf k}p_{\bf k}^*+v_{\bf k}v_{\bf k}^*\left(k^2-\frac{2}{\eta^2}\right)\right]\,\,
\ee
where $p_{\bf k}\equiv\delta\mathcal{L}/\delta {v_{\bf k}^*}'=v_{\bf k}'$, where $'$ is $\frac{\partial}{\partial\eta}$. This looks like the Hamiltonian of a scalar field, but with a time-dependent mass. We quantize in the Schr\"{o}dinger formalism. It has been shown to yield results identical to the operator or Heisenberg formalism, see \cite{Polarski:1995jg} for the inflationary scenario, and \cite{Hatfield} for a general discussion. Quantization can be done by promoting $v_{\bf k}$ and $p_{\bf k}$ to operators obeying  $[v_{\bf k}^R, p_{\bf q}^R]=i\,\delta({\bf k}-{\bf q})$\,and~$[v_{\bf k}^I, p_{\bf q}^I]=i\delta({\bf k}-{\bf q})$ where the superscripts indicate real and imaginary parts. There is a natural choice of initial state, which is the ground-state of the oscillators in the short-wavelength limit $|k\eta|\gg1$, known as the Bunch-Davies (BD) vacuum. Since the Fourier modes are non-interacting, the functional Schr\"{o}dinger equation $i\frac{\partial}{\partial \eta}{\bf \Psi}[v(\eta,{\bf x})]=\mathbb{H}{\bf \,\Psi}[v(\eta,{\bf x})]$ can be easily solved by the Ansatz ${\bf \Psi}[v(\eta,{\bf x})]\equiv\prod_{\bf k}\psi_{\bf k}^R\left(v_{\bf k}^R,\eta\right)\,\psi_{\bf k}^I\left(v_{\bf k}^I,\eta\right)$\,. We recover a Schr\"{o}dinger equation for each Fourier mode, 
\be\label{Schro}
\left[i\frac{\partial}{\partial\eta}+\frac{1}{2}\frac{\partial^2}{\partial (v_{\bf k}^R)^2}-\frac{1}{2}\left( k^2-\frac{2}{\eta^2}
\right)\left(v_{\bf k}^R\right)^2\right]\psi_{\bf k}^R=0\,
\ee
and an identical one for $v_{\bf k}^I$. We solve with the Ansatz, $\Psi_{\bf k}(v_{\bf k},\eta)=N_{k}(\eta)~exp(-\Omega_k(\eta)^2 |v_{\bf k}|^2)$. 
Applying the BD initial condition, $\Omega_{\bf k}(k\eta\rightarrow-\infty)=\frac{k}{2}\,$, we find
\be\label{solution}
\Omega_{\bf k}=\frac{k}{2}~\frac{(k\eta)^2+i/(k\eta)}{(k\eta)^2+1}\,.
\ee
The largest length-scales observable on the CMB today correspond to $|k\eta|\ll1$,
\be\label{asymp}
|\Psi_{\bf k}(\zeta_{\bf k})|^2\rightarrow \frac{\,k^{3}}{\pi H^{2}}exp\left[-k^3|\zeta_{\bf k}|^2/H^2\,\right]\,,
\ee
It is a time-independent solution of the harmonic oscillator equation $i\frac{\partial}{\partial\eta}{\Psi_{\bf k}}=-\frac{1}{2}\frac{\partial^2}{\partial|\zeta_{\bf k}|^2}\Psi_{\bf k}+\frac{k^6}{2H^4}|\zeta_{\bf k}|^2\Psi_{\bf k}$.
We can easily compute the two-point correlation function, \small
\be\label{twopt}
\langle{\bf \Psi}|\zeta_{\bf k}\zeta_{\bf p}^*|{\bf \Psi}\rangle=\frac{2\pi^2}{k^3}\mathcal{P}_\zeta(k)\,\delta({\bf k}-{\bf p})~;~~\mathcal{P}_\zeta(k)=\left(\frac{H}{2\pi}\right)^{2}.
\ee\normalsize
The delta function comes from the assumption of non-interaction of Fourier modes; any two modes are uncorrelated. The random field $\zeta({\bf x})$ is completely characterized by this two-point function. $\mathcal{P}_{\zeta}$ being independent of $k$ is the renowned Harrison-Zel'dovich scale-invariant power-spectrum.  

The inflationary paradigm provides a profound quantum mechanical origin of $\zeta(x)$, and completes the story of LSS formation \cite{Peebles}. However, a deep question remains: how exactly did quantum fluctuations during inflation become classical perturbations? In Weinberg's words, ``the field configuration must become locked into one of an ensemble of classical configurations... It is not apparent just how this happens...'' \cite[pg. 476]{weinberg2008}. Symbolically, ${\bf \Psi}[\zeta({\bf x})]\xrightarrow{?} \zeta({\bf x})$, where quantum fluctuations are described by a wave-functional ${\bf \Psi}[ \zeta({\bf x)}]$ over classical configurations $\zeta({\bf x})$. 

There is discussion in the literature of a so-called quantum-to-classical transition during inflation \cite{Starobinsky:1980te,Guth:1985ya,Albrecht:1992kf}.
 The quantum state of each Fourier mode $v_{\bf k}^R$ become ``squeezed.'' The two-dimensional Wigner function, a generalization of a classical probability distribution in phase space to quantum variables, becomes elongated in one direction, becoming a cigar-like shape. The bivariate probability distribution effectively reduces to a probability distribution of a single variable. The squeezing happens approximately in the direction of the canonical momentum operator $p_{\bf k}$ \cite{Martin:2012pea}; in this sense, it approximately becomes a c-number, and hence commutes with $v_{\bf k}$. It is easy to show this by computing expectation values \cite{Albrecht:1992kf},
\small
\be
\langle{v}^{R}_{\bf k}{p}_{\bf k}^{R}\,\rangle=\frac{i}{2}\left(1+i\,\frac{Im\,\Omega_{\bf k}}{Re\,\Omega_{\bf k}}\right);~\langle{p}^R_{\bf k}{v}^R_{\bf k}\rangle=\frac{i}{2}\left(1+i\,\frac{Im\,\Omega_{\bf k}}{Re\,\Omega_{\bf k}}\right)-i \nonumber
\ee
\normalsize
When  the physical wavelength $\lambda_{phys}\equiv{ 2\pi a(\eta)}/k$  becomes much larger than $1/H$, that is, when $|k\eta|\ll1$,  we have $Im\,\Omega_{\bf k}/Re\,\Omega_{\bf k}\simeq \left(1/k\eta\right)^3\gg 1$. That is, the expectation value of each product becomes large compared to the commutator. 

However, even though the commutator becomes small in this sense, $v_{\bf k}$ is not  a c-number. It is {\it assumed} that $v_{\bf k}$ somehow becomes a c-number after leaving the horizon \cite{lyth2009primordial,padmanabhan:1996}. The  justification provided is that, when $\lambda_{phys}$ becomes larger than the causal length-scale, which is of order $1/H$,  its physics should cease and the mode should ``freeze." But technically, we are replacing an operator by a c-number, which is akin to making a measurement in standard quantum mechanics. It is further assumed that, after this measurement, $v_{\bf k}$ remains frozen until it re-enters the horizon during the radiation era after inflation, upon which it begins to oscillate classically, and subsequent evolution of the mode is purely classical. 

The Wigner function is also connected to the density matrix, whose off-diagonal elements can vanish due to various interactions with environments, known as decoherence \cite{Campo:2005sy,Burgess:2006jn, Blencowe:2012mp}. While decoherence proposals are interesting in that they effectively ``erase'' various quantum correlations, they do not solve the measurement problem. 
\\
\\
\noindent{\bf Measurement Problem In Quantum Mechanics -- }Let us do a brief review; the basic idea and notation will carry over to the cosmological case. Consider
\be
|\psi\rangle=\mathbb{I}|\psi\rangle=\int dx~\vert x\rangle\langle x\vert\psi\rangle=\int dx~c_x \vert x\rangle\,, \nonumber
\ee
the quantum state of a particle. We inserted a completeness relation and  defined the coefficient $c_x\equiv\langle x\vert\psi\rangle$. For example, the particle could be prepared in the laboratory to be in a Gaussian state, $c_x=e^{-x^2/2}/\pi^{1/4}$. The particle is in a superposition of all position eigenstates $\{\vert x\rangle\}$. If a measurement of the particle's location is made (e.g. by shining a laser), then in the standard Copenhagen interpretation the state  is said to  {\it collapse,} 
\be\label{qm}
\boxed{
|\psi\rangle=\int dx~c_x \vert x\rangle~~{\bf \xrightarrow{?}}~~\vert x^{col}\rangle\,,\nonumber
}
\ee
where $x^{col}$ is the outcome of the measurement. This is said to happen instantaneously, and so is distinct from smooth, unitary time-evolution.  Collapse of the wavefunction is taken as a postulate  -- quantum mechanics does not explain how it happens, \emph{i.e.} its dynamics. 
\\
\\
\noindent{\bf Cosmological Measurement Problem -- }Consider a Fock space expansion of our field $\zeta(x)$ in the field-amplitude basis. $\vert{\bf \Psi}\rangle=\bigotimes_{\bf k}\, \vert\Psi_{\bf k}\rangle$. Using $\mathbb{I}=\bigotimes_{\bf k}\int{d\zeta_{\bf k}}|\zeta_{\bf k}\rangle \langle\zeta_{\bf k}|$, and defining $c_{\zeta_{\bf k}}\equiv\langle\zeta_{\bf k}|\Psi_{\bf k}\rangle$ and
\ba
\int\prod_{\bf k}d\zeta_{\bf k}\equiv\int\mathcal{D}[\zeta]~;~~\prod_{\bf k}c_{\zeta_{\bf k}}\equiv c[\zeta]~;~~\bigotimes_{\bf k}|\zeta_{\bf k}\rangle\equiv\vert[\zeta]\rangle\,,\nonumber
\ea
where $[\zeta]$ is a specific configuration of the field in Fourier space, and $\int\mathcal{D}[\zeta]$ represents integration over all such configurations, we can  express the Problem formally as,
\be\label{CMP}\boxed{
\vert{\bf \Psi}\rangle=\int~\mathcal{D[}\zeta]~~c[\zeta]~~\vert[\zeta]\rangle ~~~~{\bf \xrightarrow{?}} ~~~~\vert[\zeta]^{col}\rangle\,,}
\ee
where $[\zeta]^{col}$ is the collapsed configuration. We  already calculated the coefficients $c[\zeta]$ in \eqref{asymp}, $|c_{\zeta_{\bf k}}|^2=|\Psi_{\bf k}(\zeta_{\bf k})|^2$. 

Equation \eqref{CMP} shows explicitly that during inflation, the curvature perturbation field is in a linear superposition of all possible field configurations $\{\vert[\zeta]\rangle\}$. 
We believe that the collapse  happened in this field-amplitude basis because it is the pointer basis for cosmology \cite{pointer}; the field-amplitude operator $\zeta_{\bf k}$ commutes with standard interaction Hamiltonians such as $\zeta^4$ and $\zeta^2\chi^2$. This means that, once the collapse to some field-amplitude eigenstate  has taken place, further interaction of the field with itself and any environment will preserve the eigenstate. 

We are faced with two logical possibilities. Either there is an issue with Copenhagen interpretation -- there were obviously no observers in the early Universe who could have made the measurement, as any observer, such as ourselves, would owe their existence to the primordial density perturbations -- or there is some unknown dynamics that ``selects" one field configuration $[\zeta]^{col}$.  Let us discuss what this  dynamics could have been.\\
\\
{\bf A Solution -- }We begin with the discrete case. The Fourier transform is defined as 
$v({\bf x},\eta)=\sum_{\bf k} \,v_{\bf k}(\eta)\,e^{-i{\bf k}\cdot{\bf x}}\,$, with ``grid-spacing" $k_s$, that is, $|{\bf k}-{\bf q}|\geq k_{s}$ for distinct modes ${\bf k}$ and ${\bf q}$. We will discuss the continuum limit $k_s\rightarrow0$ shortly. Instead of the Hamiltonian \eqref{Ham}, we have $ \mathbb{H}=\mathbb{H}_0+\mathbb{H}_{int}$, where $\mathbb{H}_0=\sum_{\bf k} \left[p_{\bf k}p_{\bf k}^*+v_{\bf k}v_{\bf k}^*\left(k^2-\frac{2}{\eta^2}\right)\right]$ and we propose the following two-body interaction between modes,
\be\label{interaction} 
\mathbb{H}_{int}~=~\frac{1}{2\pi}~~\sum_{\bf k}~\sum_{{\bf q}\,:~|{\bf q}- {\bf k}|<\Delta}~\gamma\,(k,\eta)~~~\delta(R_{\bf k}-R_{{\bf q}})\,,
\ee
where we have absolute value $R_{\bf k}\equiv~|\zeta_{\bf k}|$,  coupling strength $\gamma$, and ``interaction window" $\Delta$.
\vspace{0.25cm}
\begin{center}
\includegraphics[scale=.4]{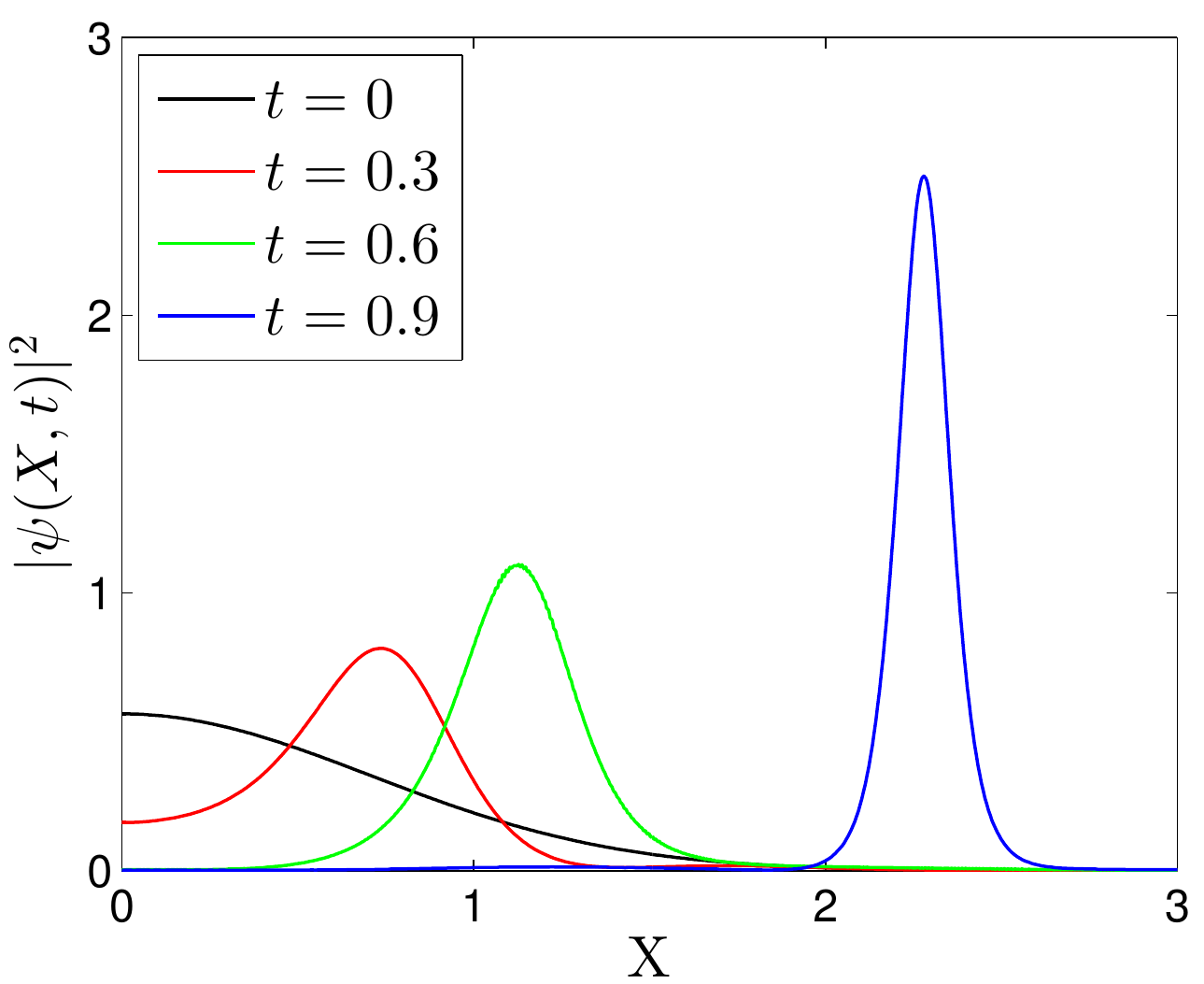}
\end{center}
\captionof{figure}{(Color online) Numerical solution of   \eqref{adim}, with $\Gamma=-20$. For simplicity, we used co-moving time $t$ here instead of $T$. The effective non-linear term modifies the traditional Gaussian profile $|\psi|^2$ (Black), into an arbitrarily sharp spike (Blue). We found the shape of the spike to be asymptotically Gaussian. See Fig. 2 for the full wavefunction $|\Psi|^2$\,.}
\vspace{0.25cm}
\hspace{0.25cm} Our interaction is inspired by the renowned ``hard-sphere" model in a two-dimensional weakly-interacting Bose gas \cite{Huang}, with  identification of  ${\bf k}$ and $\zeta_{\bf k}$ to a Bose particle with label $j$ and position ${\bf r}_{j}$, respectively. It is easy to check  consistency; $\zeta_{\bf k}$ and ${\bf r}_j$ have the same units (setting mass of Bose particle to unity).

\hspace{0.25cm} Since there is explicit $k$-dependence in our Hamiltonian (as if there was label-dependence in a Bose gas), we need to assume $k_s\ll\Delta\ll k$. The physical meaning of this is that, every  ``particle" ({\it i.e.} Fourier mode) {\it within that window} is equivalent, and every two-body interaction between those pairs of particles is, on average, equivalent (Hartree-Fock approximation). That is, we assume 
\be\label{HF}
{\bf \Psi}(\zeta_{\bf k_1},\zeta_{\bf k_2},...,\zeta_{\bf k_N},\eta)\simeq  ~\left[\Psi\left(\zeta_{\bf k},\eta\right)\right]^N\,,
\ee
where $N$, the number of modes in that window, scales as $(\Delta/k_s)^3$. We will further assume, 
\be\label{sepvar}
\Psi(\zeta_{\bf k},\eta)\equiv\psi(R_{\bf k})\,\Theta(\theta_{\bf k})\,,
\ee and integrate out the angular part. This is valid since the real and imaginary parts of $\zeta_{\bf k}$ can be quantized independently as described before, and the BD initial condition is such that there is no angular dependence. We have the normalization condition $\int_0^\infty dR_{\bf k}~R_{\bf k}\int_0^{2\pi}d\theta_{\bf k}~|\Psi\left(\zeta_{\bf k},\eta)\right|^2=1$.

\hspace{0.25cm} It has been shown \cite{Lieb2014} that the HF approximation \eqref{HF} is valid for a two-dimensional Bose gas with any radially-symmetric interaction in the limit of large $N$ and  correspondingly small coupling $\gamma$, just as it is for a three-dimensional Bose gas. Physically, this limit means that any two given modes are  weakly coupled, {\it i.e.} they are uncorrelated to a good approximation.

\hspace{0.25cm} Under these assumptions, we recover a  Schr\"{o}dinger equation with a new effective non-linear term, $i\frac{\partial}{\partial\eta}\psi(R_{\bf k},\eta)=-\frac{1}{2}\frac{\partial^2\psi}{\partial(R_{\bf k})^2}+\frac{k^6R_{\bf k}^2}{2H^4}\,\psi+N\gamma\,R_{\bf k}\,|\psi|^2\psi\,.$ We assume our interaction is negligible at the beginning of inflation, by appropriate choice of $\gamma$ \footnote{We are imagining that the interaction gets switched on near the end of inflation. For example, the interaction could depend on the time-derivative of the deceleration parameter $\dot{\epsilon}$, and so is triggered when the inflaton reaches the bottom of its potential and begins to oscillate.}. This means the BD vacuum with Gaussian wavefunction is still the natural initial state, but the non-linear term becomes important after a few e-folds. 
Upon a-dimensionalizing,
\be\label{adim}
i\frac{\partial}{\partial T}\psi(X,T)=-\frac{1}{2}\frac{\partial^2\psi}{\partial X^2}+\frac{1}{2}X^2\psi+\Gamma\,X|\psi|^2\psi\,,
\ee
we have the dimensionless variables 
\be
X\equiv R_{\bf k}\,k^{3/2}/H\,,
\ee
and $T\equiv \eta~k^3/H^2$, and the lone free parameter $\Gamma\equiv\gamma\,N\,k^{9/2}/H^3$. (Fig. 1). The location of the peak for $|\psi(R_{\bf k})|^2$ needs to scale as $1/k^{3/2}$ in order to have a scale-invariant power spectrum.
This can be achieved by adjusting the $k$-dependence and time-dependence of $\gamma$, as follows. We set $\gamma\propto k^{-9/2}$, and  the time dependence of $\gamma$ such that the length of time our interaction term is switched on is proportional to $k^3 /H^2$. This choice of behavior of $\gamma$ would effectively make $\psi(X,\eta)$  independent of $k$, and hence $\vert\psi(R_{\bf k},\eta)\vert^2$ will have the desired $1/k^{3/2}$  scaling. Finally, since $\theta_{\bf k}$ is a random variable uniformly distributed over $[0 ~2\pi)$, we will recover the classical Gaussianity of $a_{lm}$'s via CLT. The amplitude and duration of $\gamma$  also controls the precision of  collapse.
\\
\\
{\bf Discussion -- }Let us elaborate on \eqref{interaction}. Since the HF approximation  is valid in the  $N\rightarrow\infty$ limit, we may consider taking $k_s\rightarrow0$ and replacing $\sum_{\bf k}$ with $\int d^3k$\,, \small
\be
\mathbb{H}_{int}=\int d^3k~\int_{k-\Delta}^{k+\Delta} d^3q~\bar{\gamma}~\delta(R_{\bf k}-R_{{\bf q}})=\int d^3k~\frac{\bar{\gamma}\,}{|\nabla_{\bf k}R_{\bf k}|} \nonumber
\ee \normalsize
where $\bar{\gamma}$ is some modification of $\gamma$ for the continuum limit.
It is easy to see that this is distinct from traditional $\lambda\phi^4$ types of interactions. One way is to write the delta function as a limiting case of a narrow Gaussian, and Taylor expand. The term $R_{\bf k}R_{{\bf q}}$ and all higher powers will be non-negligible, unlike a $\lambda\phi^4$ theory. In addition, our theory is non-local; consider the Schwinger expansion,\small
\be
\frac{1}{|R_{\bf k}'|}=\int_0^1 e^{-\xi\,|R_{\bf k}'|}~d\xi
=\int_0^1 d\xi\left[1-\xi|R_{\bf k}'|+\frac{1}{2}\xi^2|R_{\bf k}'|^2+...\right]\nonumber
\ee
\normalsize
where $'$ is short-hand for $\nabla_{\bf k}$. This formula allows us to express the Hamiltonian in real space; the terms look like $\mathbb{H}_{int}=\int d^3x~R(x)\int d^3\tilde{x}~\tilde{x}^2R(\tilde{x})^2 + ...$~. Interestingly, in standard quantum field theories, position is a label and Lagrangians are usually label-free. Ours has label dependence; we may call it  the ``eye of God.'' Further investigation into its physical meaning is ongoing. \cite{ultimo} 
\\
\\
\noindent {\bf Conclusion -- }Thanks to CLT, our interaction Hamiltonian can reproduce the standard Gaussian predictions for $a_{lm}$, while attempting a solution of the Measurement problem in Inflation (Fig. 2). This Problem and our solution is relevant for other scenarios in cosmology as well, such as cyclic or bounce universes, and modified dispersion relation theories \cite{Arzano:2015gda}.
\vspace{0.25cm}
\begin{center}
  \includegraphics[scale=.35]{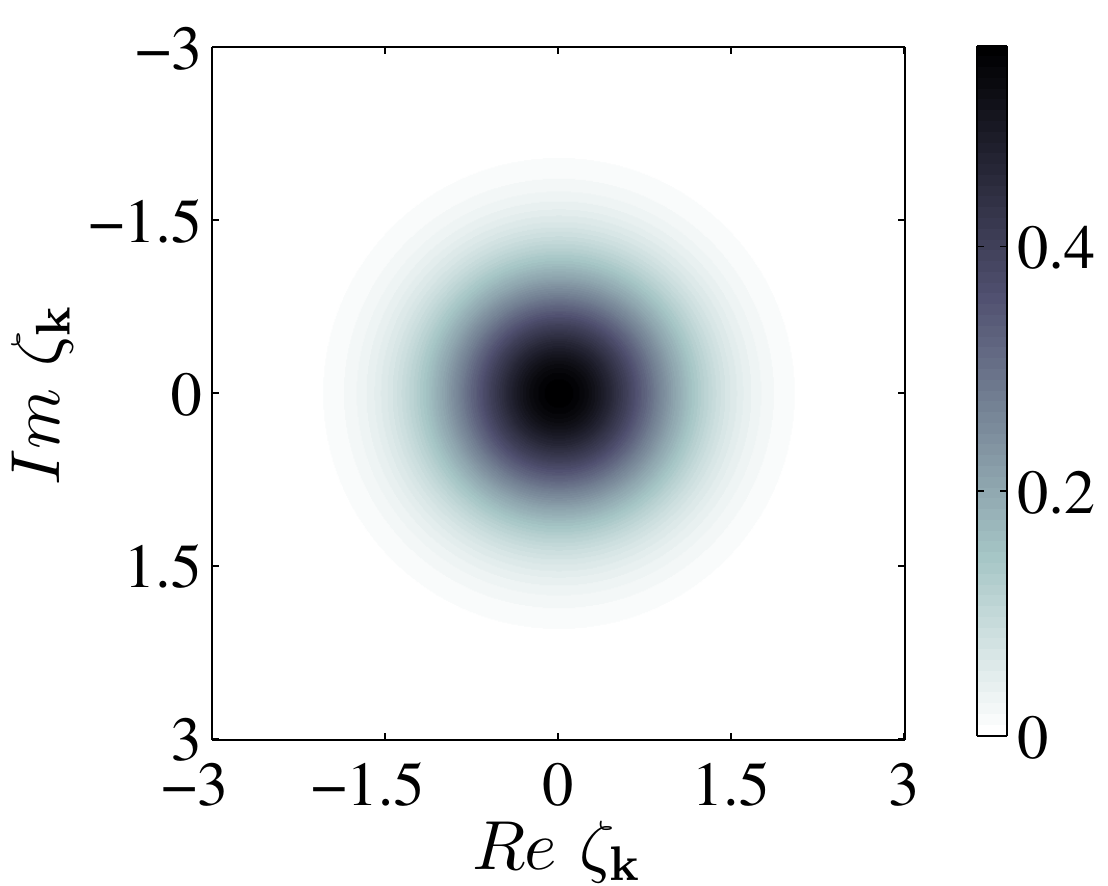}\hspace{.5cm}\includegraphics[scale=.35]{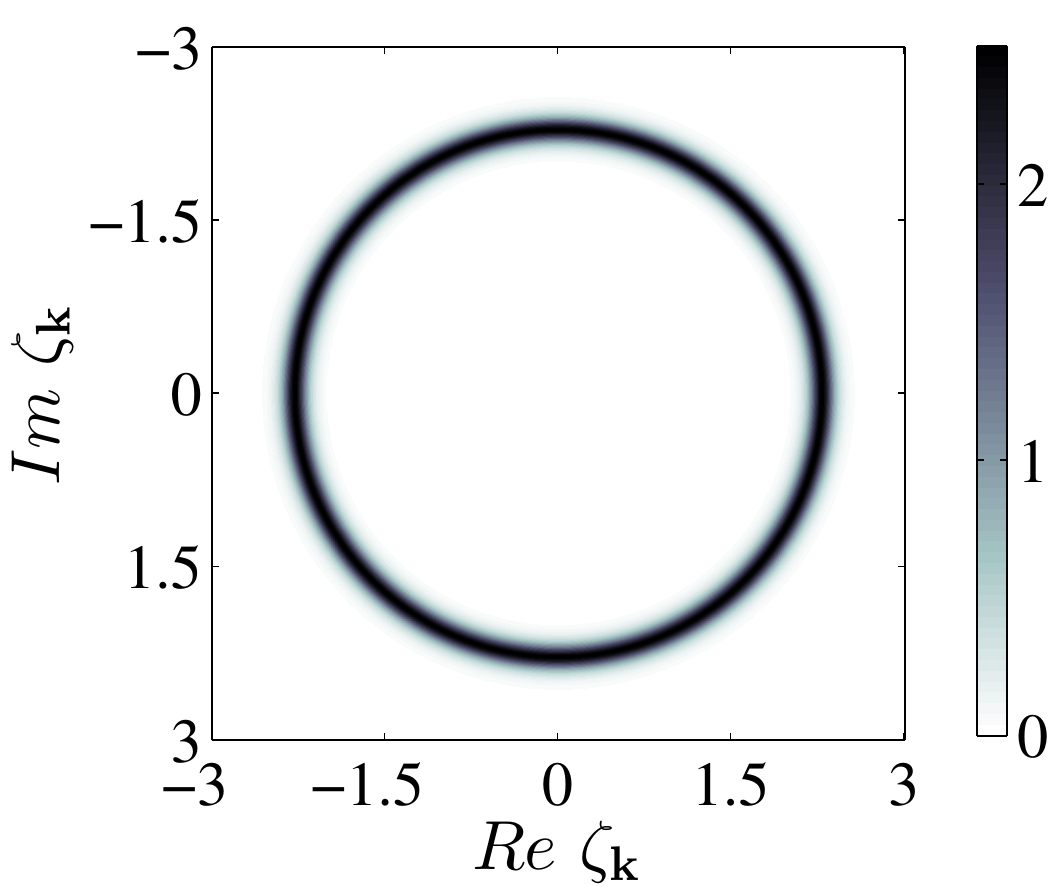}
\end{center}
\captionof{figure}{The initial (BD) and final wavefunction $|\Psi(\zeta_{\bf k})|^2$\,, setting $k^{3/2}/H=1$. Essentially, we reduced the dimensionality of the wavefunction manifold from two (amplitude and phase, $R_{\bf k}$ and $\theta_{\bf k}$, respectively) to one (just $\theta_{\bf k}$). This remaining stochasticity of $\theta_{\bf k}$ is less disturbing than that of $R_{\bf k}$ implicit in the traditional description.}
\vspace{0.25cm}
\hspace{0.25cm} $\theta_{\bf k}$ is responsible for the transition from the perfect rotational and translational symmetry of the  quantum  state (BD vacuum) to the (slightly)  inhomogeneous and anisotropic universe that we live in \footnote{We are grateful to Prof. D. Sudarsky for emphasizing this point (private communication).}. It is the complex phase of $\zeta_{\bf k}$; it corresponds to translations in physical space, and is a random variable both in the standard description \cite{Albrecht:1992kf} and in our model. It is distinct from the phase of standing waves that lead to Sakharov oscillations in the CMB, discussed in \cite{Dodelson:2003ip}.  It will be interesting to further investigate the origin and behavior of  $\theta_{\bf k}$,  which we shall address elsewhere.

\hspace{0.25cm} The cosmological measurement problem is a rich and compelling arena for both foundational issues of quantum mechanics as well as a deep understanding of early universe cosmology, and may potentially teach us about aspects of quantum gravity.
\\
\\
\noindent{\bf Acknowledgements -- }We are grateful to Walter Lawrence III, Tirthabir Biswas, Leon Cooper, Daniel Sudarsky, Robert Caldwell and Ward Struyve for helpful comments. DJ thanks his Dartmouth family. Work at Dartmouth was supported in part by DOE grant DE- SC0010386. JM acknowledges support from John Templeton Foundation, a STFC consolidated grant and the Leverhulme Trust. 
\bibliography{measurement}
\end{document}